\documentclass[twocolumn,showpacs,amssymb,aps,nofootinbib]{revtex4}
\usepackage{psfig,epsfig}
\newcommand{\be}{\begin{eqnarray}}
\newcommand{\ee}{\end{eqnarray}}
\newcommand{\ave}[1]{\left\langle #1 \right\rangle}

 \newcommand{\Nch}{N_{\rm ch}}
\begin{document} \hbadness=10000
\topmargin -0.8cm\oddsidemargin = -0.7cm\evensidemargin = -0.7cm
\title{Particle yield  fluctuations and chemical  non-equilibrium at RHIC}
\author{Giorgio Torrieri}
\affiliation{
Department of Physics, McGill University, Montreal, QC H3A-2T8, Canada}
\author{Sangyong Jeon}
\affiliation{
Department of Physics, McGill University, Montreal, QC H3A-2T8, Canada}
\affiliation{
RIKEN-BNL Research Center, Upton NY, 11973, USA}
\author{Johann Rafelski}%,
\affiliation{
Department of Physics, University of Arizona, Tucson AZ 85721, USA}

\date{Revised February 2006}  %Minor revisions August  2005 v 3.27

\begin{abstract}
We study charge fluctuations within the statistical hadronization
model. Considering both  the particle yield ratios and the
charge fluctuations we show that it is possible to
differentiate between  chemical equilibrium and non-equilibrium  freeze-out
conditions. As an example of the procedure we show  quantitatively how
the relative yield  ratio  $\Lambda/K^-$
together with the normalized net charge fluctuation
$v(Q)=\ave{\Delta Q^2}/\ave{\Nch}$  constrain  the chemical conditions at
freeze-out.   We also discuss the influence of the
limited detector acceptance on fluctuation measurements, and show how
this can be  accounted for within a quantitative analysis.
\end{abstract}
\preprint{CERN-PH-TH/2005-041}
\maketitle
%%%%%%%%%%%%%%%%%%%%%%%%%%%%%%%
\section{introduction}
In relativistic heavy ion collisions a localized high energy density
domain, a fireball, is created. The study of the properties of this
hot and dense matter is the main objective of the experiments being
conducted at RHIC and as of 2007 at LHC.
Event-by-event particle fluctuations are  the observables
 subject to  intense current
theoretical~\cite{fluct1,fluct2,fluct3,fluct4,fluct4b,
fluct5,fluct6,fluct7,fluct8},
and experimental ~\cite{starfluct,starfluct2,phefluct} interest.
Fluctuation measurements are important since they can be used:
(i) as a consistency check for existing models, e.g. within statistical
particle production models~\cite{fluct2,fluct3},
(ii) as a way to search  for new physics, including QGP
\cite{fluct4,interm2,interm3}
(iii) as a test of particle equilibration \cite{fluct2,fluct8}.

The statistical
hadronization model (SHM), introduced by Fermi in 1950
\cite{Fer50,Pom51,Lan53}, has been used extensively in recent years
in the study of strongly interacting particle production. In this model,
the properties of the final state particles are determined by requiring that
the final state maximizes entropy given the physical properties of the
fireball (energy, baryon  content, etc.).
 When the full spectrum of hadronic
resonances is included  \cite{Hag65}, the SHM
turns into a quantitative model
capable of describing
in detail
the abundances of all hadronic particles.

Fluctuations in conserved quantum numbers (such as charge,
baryon number, strangeness, or
equivalently the net multiplicities of
up, down and strange quarks)
can be studied only
in the Grand Canonical (GC) ensemble,
since
in the micro-canonical and
canonical  ensembles these quantities are fixed.
We also mention here
that fluctuations of non-conserved observables, e.g. other
hadron multiplicities, differ for different
ensembles even in the thermodynamic limit \cite{nogc1,nogc2}.

In this paper we will discuss the use of fluctuations as a phenomenological
tool within the framework of the statistical model, and illustrate some issues
pertinent in analyzing
fluctuations data.   In section \ref{obschoice}
we will motivate the choice of charge fluctuations
as a useful experimental probe.  After demonstrating, in section \ref{secshm},
how the statistical model implies a scaling between
fluctuations and yields, we show (section \ref{noneq}) that a measurement
of both particle yields and charge fluctuations can distinguish between
an equilibrium high temperature statistical freeze-out from a super-cooled
over-saturated freeze-out from a high entropy phase.
Finally, in section \ref{secacceptance} we discuss issues related to detector
acceptance which impact the fluctuation measurement even in a boost-invariant
azimuthally symmetric limit.   We quantitatively demonstrate how such limited
acceptance effects can be taken into account and the freeze-out temperature
and non-equilibrium parameters extracted from experimental data.

\section{\label{obschoice} GC Observables}

A study of GC SHM fluctuations of conserved
quantities is of considerable interest at RHIC.
Since the detectors at RHIC (except for the PHOBOS detector) only
see small portions of the final phase space,
using the grand-canonical approach is justified in the following sense:
Provided the fireball is indeed locally thermalized,
we can take
the experimentally observed source to be a subsystem in contact with a
larger reservoir.
%
%domain exchanges particles
%and energy with the unobserved much larger background
%``bath'', giving rise to fluctuations in conserved quantum numbers, in the
%same sense that the density of air molecules fluctuates locally even though
%their number is conserved (no particle production).

The situation is of particular interest for
reactions at RHIC that exhibit a sizable
central plateau in the (pseudo-)rapidity spectrum,
since
a limited (pseudo)rapidity acceptance window
selects  a suitable subset of the
source particles. Specifically, 
it can be shown (sections \cite{cleymans}.  The reasoning used there can be generalized to Fermi-Dirac and Bose-Einstein statistics) that
the rapidity spectrum of a boost invariant system
%in a window around mid-rapidity
could be related to the multiplicity in a static GC system with
the same temperature and chemical potentials
\be
{\ave{dN_i/dy}_{\rm b.i.} \over \ave{dN_j/dy}_{\rm b.i.}}
=
{\ave{N_i}_{\rm GC}\over \ave{N_j}_{\rm GC}}
\label{eq:boost_inv}
\ee
where $i$ and $j$ are species labels and the subscripts ${\rm b.i.}$ and
${\rm GC}$ denote the boost invariant system and the grand canonical
system, respectively.

The derivation in \cite{cleymans} can be applied to fluctuations at hadronization (before resonance decays) to show
\be
{\ave{d{\Delta N_i^2}/dy}_{\rm b.i.} \over \ave{dN_j/dy}_{\rm b.i.}}
=
{\ave{\Delta N_i^2}_{\rm GC}\over \ave{N_j}_{\rm GC}}
\ee
where we denote the variance (fluctuation) of any quantity $X$ as
$\ave{\Delta X^2} = \ave{X^2}{-}\ave{X}^2$.

Given this, SHM average yields and yield fluctuations can be calculated by a textbook
method \cite{huang}, as per section
\ref{secshm}.

When studying finite systems  the  consideration
of fluctuations in {\em extensive} quantities such
as of particle yield has to address also
volume fluctuations when the volume cannot be fixed
by experimental conditions.
In our case volume fluctuations
can arise due to initial reaction
effects, impact parameter variations, as well as from
fluctuations due to dynamics of the expanding fireball.
It is difficult to arrive at a reliable description of all these effects.
Therefore it is important to select fluctuation  observables in which
volume fluctuation effects are sub-dominant.
 Among extensive quantities,
the net charge fluctuation stands out as it is relatively easy to measure
and can be shown to be nearly independent of the volume fluctuations
\cite{fluct1}.

In light of the above considerations
we concentrate our effort on the
following net charge fluctuation measure:
\be
\label{vqdef}
v(Q) \equiv \left< \Delta Q^2 \right>/\left<\Nch\right>
\ee
(where $N_{\rm ch}=N_+ + N_-$)
proposed in the past as a probe of the QGP formation
\cite{fluct4}. First results   for $v(Q)$ are also available from RHIC
experiments
\cite{starfluct2,phefluct}.

In the SHM,
the charged particle multiplicity is given by summing
all final state (stable) charged particle multiplicities.
These can be computed by adding the direct
yield and all resonance decay feed-downs.
The total yield of a stable particle $\alpha$ is
\be
 \label{resoyield}
\langle N_\alpha\rangle_{\rm total} & = &
\langle N_\alpha\rangle_{\rm GC} + \sum_{j\ne \alpha}
B_{j \rightarrow \alpha}  \langle   N_j \rangle_{\rm GC}
 \label{fluctdef}
 \ee
where $j$ labels resonances.
$B_{j \rightarrow \alpha}$ is the probability (branching ratio)
for the decay products of $j$ to include $\alpha$.
The charged particle multiplicity is given by the sum of all
charged stable particles.

The net charge fluctuation is given by
\be
\ave{\Delta Q^2}_{\rm GC} = \sum_{i} q_i^2 \ave{\Delta N_i^2}_{\rm GC}
\label{eq:DQ2}
\ee
where $q_i$ is the particle charge and
$i$ labels {\em all} particles {\em before} resonance
decays since  net charge is conserved \cite{fluct3}.

To use Eq.(\ref{eq:DQ2}) quantitatively,
however, the experimental rapidity window
must be large enough to encompass all
decay particles of the resonances, yet small enough for the GC ensemble to
maintain it's validity.   See section \ref{secacceptance} for a discussion of
the validity of this assumption, and how to incorporate deviations from it in
realistic experimental
situations.

\section{\label{secshm}Statistical hadronization}

For a hadron with an energy
$E_{p} = \sqrt{p^2+m^2}$, the GC partition function for each species is
given by
\begin{eqnarray}
\label{partition_function}
\ln Z_i =
(\mp) V g_i\int {d^3p\over (2\pi)^3}
\ln \left(1 \pm \lambda_i e^{-E_i/T}  \right)
\end{eqnarray}
where
$g_i$ is the degeneracy factor and the upper sign is for bosons and
the lower sign is for fermions.
Here $\lambda_i$ is the particle fugacity, related to particle chemical
potential $\mu_i=T\ln \lambda_i$.

The yield average and fluctuation is then given by:
\be
\label{yield_formula}
\langle N_i\rangle_{\rm GC}
& = & \frac{\partial \ln Z_i }{\partial \lambda_i}
  =  g_iV\int  {4 \pi p^2 dp \over (2\pi)^3}\, n_{i}(E_p),
\\
\label{fluct_formula}
\ave{\Delta N_i^2}_{\rm GC}
& =& \frac{\partial^2 \ln Z_i }{\partial \lambda_i^2} \nonumber \\  &=&
g_iV\int {4 \pi p^2 dp \over (2\pi)^3}\, n_{i}(E_p) \left(1 \mp
n_i(E_p)\right).
\end{eqnarray}
and
\begin{equation}
n_{i}(E_p) = {1\over  \lambda_i^{-1} e^{E_p\beta}\pm 1},
\end{equation}
%Eqs.\,(\ref{yield_formula}--\ref{fluct_formula}) can be
%evaluated to any desired accuracy
% in terms of a series of   Bessel functions~\cite{Lan53} when $m_i>\mu_i$.
%%Otherwise,
% numerical evaluation is required -- this occurs for degenerate Fermi gases.

We note that $\lambda_i$ enters   the {\em partition function} in
Eq.(\ref{partition_function}).   Hence, the validity of
Eqs.(\ref{yield_formula})
and (\ref{fluct_formula}) depends on weather
Eq.(\ref{partition_function}) can be
used as a {\em generating function} for the probability distribution of states.
  It is important to underline this as in a dynamical system the value of $\lambda_i$ is
not determined solely in terms of entropy maximization, but is subject
to chemical conditions prevailing in the system, and here importantly, includes
effects related to  chemical  non-equilibrium.
Where Eq. \ref{partition_function} represents a generating function but the system is not in chemical equilibrium, the fugacity  $\lambda_i$, is not anymore a Lagrange multiplier but a parameter
characterizing the quantum number density.

In a scenario where freeze-out occurs as a break-up of a
{\em chemically equilibrated} hadron gas,
the fugacity of the hadron $i$ is given
by the product of the fugacities of conserved quantum numbers.
\begin{equation}
\lambda_i^{\mathrm{eq}} = \lambda_{q}^{q-\overline{q}} \lambda_{s}^{s -
\overline{s}}
\lambda_{I_3}^{I_3} \;, \; \lambda_{\overline{i}}^{\mathrm{eq}} =
(\lambda_i^{\mathrm{eq}})^{-1},
\label{eqlam}
\end{equation}
where $\overline{q},q$ is the number of light anti-quarks and quarks,
respectively and
$\overline{s},s$ is the number of strange anti-quarks and quarks,
respectively and $I_3$ is the isospin.
This formula implies that the fugacity for the
antiparticle is, in full chemical equilibrium,
the inverse of the fugacity for the particle,
and the fugacity for a hadron carrying vanishing
conserved quantum numbers is 1.
%Each fugacity factor, of course, is related to a chemical potential ($
%\lambda_a = e^{\beta\mu_a}$, where $a$ can be $q, s$ or $I_3$).

In our approach, we do not assume that that the chemical equilibrium is
reached~\cite{JJBook,Rafelski:2003ju}.
Hence Eq.(\ref{eqlam}) no longer applies.
The deviation from chemical equilibrium can be
parametrized by a phase space occupancy factor
$\gamma_q$ (for $u,\bar{u}, d, \bar{d}$
in hadrons)
and $\gamma_s$ (for $s$ and $\bar{s}$).
In this {\em chemical nonequilibrium} case the fugacity becomes
\begin{equation}
\label{chemneq}
\lambda_i = \lambda_i^{\mathrm{eq}}
\gamma_q^{q+\overline{q}} \gamma_s^{s+\overline{s}}
\end{equation}
where $\lambda_i^{\mathrm{eq}}$ is given by Eq.(\ref{eqlam})
(Note that $\gamma_i = \gamma_{\overline{i}}$).

A system undergoing collective expansion is unlikely to be in chemical
equilibrium, since collective expansion and cooling will make it impossible for
endothermic and exothermic reactions, or for creation and destruction
reactions of a rare particle, to be balanced.   However, since inelastic
collisions have in general a slower relaxation time than elastic ones, an
approximately
perfect fluid can still have $\gamma \ne 1$  (it's evolution will be a
non-trivial function of time, since $\gamma$ does not commute with the
Hamiltonian).
Furthermore, light quark chemical nonequilibrium is well motivated in a
scenario
where an entropy rich deconfined state quickly hadronizes
\cite{Rafelski:2000by}.
In this scenario, mismatch of entropies between the two phases requires
$\gamma_q>1$.

Despite the lack of equilibrium and entropy maximization w.r.t. conserved
quantum numbers, we will argue that the Eqs. \ref{fluct_formula} and
\ref{yield_formula}
apply in such a situation, with $\gamma$ s contributing to the chemical
potential via Eq.(\ref{chemneq}).
The validity of Eq.(\ref{fluct_formula})
and (\ref{yield_formula}) depend on the extent that
Eq.(\ref{partition_function}) represents a probability generating
function for the statistically hadronizing system.
Within a statistical hadronization scenario where hadrons are
formed in proportion to their phase space
weight given (not necessarily equilibrated) densities
\cite{Danos}, this is indeed the case provided  the dynamics
behind $\gamma$ does not generate
additional, non-statistical fluctuations.
For an instance where the last issue is a concern, fluctuations of a quantum number produced
mostly in initial-state processes (such as charm
\cite{thews,becattini_charm}) will likely be dominated not by
the statistical hadronization contribution but to fluctuations in initial
abundance.

Given that in the considered model non-equilibrium arises due to the rapid
hadronization of the collectively expanding system \cite{Rafelski:2000by}, and
since the observable charged particles are produced not in in the initial state
but during the final break-up of a locally thermalized system, such
non-statistical fluctuations should not be significant for the observable we
are considering.   Similarly, as we have argued in the previous section,
initial-state
volume fluctuations give a negligible contribution to the observable under
consideration.

However,  it is possible that additional sources of irreducible two-particle
correlations and fluctuations
could arise near a phase transition.
These effects go beyond the scope of this work.   We will however argue that
the applicability of our scenario, and the absence of further correlations can
be {\em tested} by {\em requiring} that the same temperature and $\gamma$ s
describe both the yields and the fluctuations of {\em all} soft hadronic
observables.
As we will show, this is a very stringent requirement.
If it turns out that a single set of $T$, $\lambda^{\rm eq}$ and 
$\gamma_q$ and $\gamma_s$ is capable of describing all yields and
fluctuations, 
then it certainly is a strong indication that Eq.(\ref{partition_function})
can be interpreted as a generating function of the probabilities.
The goal of this paper is then to find a way to experimentally determine the additional
parameter $\gamma_q$  which can be then used to compare
the SHM calculation of yields and fluctuations to the experimental measurements.

\section{\label{noneq}fluctuations in chemical non-equilibrium}

Chemical nonequilibrium is of a particular interest since
it can result  in a large pion fugacity which influences fluctuations much
more severely than the yields.

If $\gamma_q$ becomes large enough so that $\lambda_{\pi}$
approaches $e^{m_\pi/T}$, then the pion yield and the fluctuations
behave like
(c.f.~Eqs.(\ref{yield_formula},\ref{fluct_formula}))
\begin{equation}
\label{divergence}
\lim_{\epsilon\rightarrow 0} \langle N\rangle \propto \epsilon^{-1},
\qquad
 \lim_{\epsilon\rightarrow 0} (\Delta N)^2 \propto \epsilon^{-2}.
\end{equation}
where $\epsilon = 1 - \lambda_\pi e^{-m_\pi/T}$. The fluctuation grows much
faster than the yield as mentioned above.

Some studies of yield ratios have indeed found the value of
$\gamma_q$ that can potentially make $\epsilon$
small~\cite{observing,Rafelski:2003ju,Rafelski:2004dp,gammaq_energy}.
However, other studies of yield ratios~\cite{Becattini:2003wp}
concluded that $\gamma_{q}$ is not necessarily large due to the fact
parameters in such fits are highly correlated.
In this case, adjusting other parameters such as
the temperature can  accommodate current
data without having $\gamma_q \ne  1$, but with much reduced statistical
significance.
Since such conflict is common when only the {\em yields} are considered,
it becomes necessary to study
fluctuations as an additional constraint
to determine the occupation factor $\gamma_q$ more convincingly.

We now discuss our specific analysis results. We used the public
domain SHM suite of programs SHARE \cite{share}, expanded to include
the fluctuations \cite{share2}. We evaluate
  yields and fluctuations, allowing for production of
hadron resonances, their decay, and
 a possible absence of chemical equilibrium.
In the rest of this paper, we set
$\lambda_{I_3}^{\rm eq}=1,
\lambda_q^{\rm eq}=e^{\mu_B/3T}=1.05$ and $\lambda_s^{\rm eq}=1.027$  in
accordance with \cite{Rafelski:2004dp}.
However, the two observables we consider, the net charge
fluctuations and the $\Lambda/{\rm K}^-$
particle yield ratio, are nearly independent of
these quantities as will be shown below.

%%%%%%%%%%%%%%%%%%%%
\begin{figure}[!tb]
\psfig{width=8.cm,clip=,figure=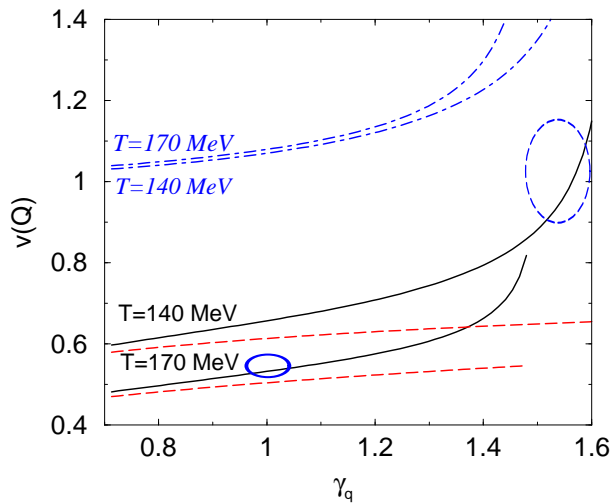}
%\vspace*{-.2cm}
\caption{(Color online)\label{datfluctboltz}
$v(Q)$ as function of $\gamma_q$ (solid lines).
Dot-dashed lines, no resonance decays;
dashed lines, Boltzmann fluctuations.
Ellipses (blue) indicate the expected result areas for
the equilibrium ($\gamma_q=1$, solid) and non-equilibrium
($\gamma_q\ne 1$, dashed) models.
%At small $\gamma_q$
%$v(Q)$ increases with $T$ for solid lines and dashed lines, decreases for
%dot-dashed lines.
%Dotted line corresponds to Poisson fluctuations.
}
\end{figure}
%%%%%%%%%%%%%%%%%%%%%%%%%%

Fig.~\ref{datfluctboltz} shows the variation in $v(Q)$ as a function of
$\gamma_q$ for $T=140, 170$ MeV. The solid lines
show  $v(Q)$ including the resonance
decays, dot-dashed lines comprise only the direct effect
of pion fluctuations.
As the temperature increases (solid lines from top to bottom)
the number of resonances increases. This in turn
increases the unlike-sign charge correlations
and hence reverses
the temperature dependence of the pure pion case (dot-dashed
lines). The short
dashed lines show results for Boltzmann statistics.
Boltzmann charge
fluctuations are nearly  constant as function of $\gamma_q$ and primarily
depend
on chemical mix  of the directly produced and secondary decay particles,
which dominantly depend on the temperature $T$.
The solid and dot-dashed  lines in Fig.~\ref{datfluctboltz}
terminate
when the fluctuations start to diverge as in
Eq.(\ref{divergence}).

\begin{figure}[!tb]
\psfig{width=8.cm,clip=,figure=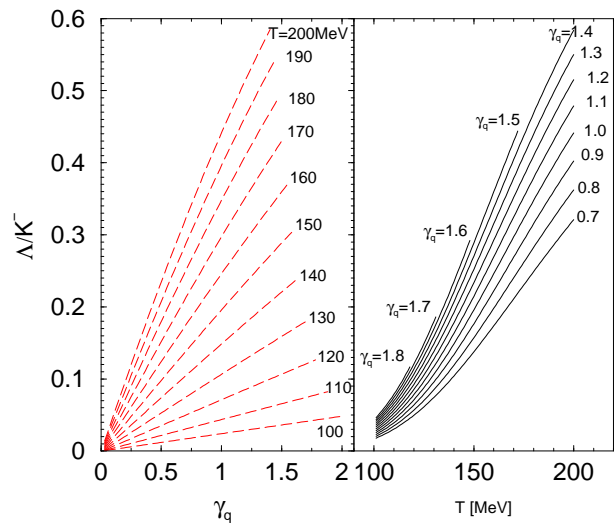}
\caption{(Color online)\label{datyld}
Particle yield ratio  $\Lambda/K^-$ as a function of $T$ (right panel) and
$\gamma_q$ (left panel)
% on left as a function of $\gamma_q$ for a range of $200<T<100$ MeV,
%and on right as function of $T$ for a range of  $1.8>\gamma_q>0.7$.
The $\Lambda$ yield does not include  $\Xi \rightarrow \Lambda$ and
the ${\rm K^-}$ yield is without the contribution of $\phi
\rightarrow {\rm K^+}{\rm K^-}$ decays.
Ellipses (blue) indicate the expected result areas for
the equilibrium ($\gamma_q=1$, solid) and non-equilibrium
($\gamma_q\ne 1$, dashed) models.
}
\end{figure}

To determine both  $T$ and $\gamma_q$ values we
  require an additional observable.
In this work, we choose the yield ratio  $\Lambda/K^-$.
This ratio  depends linearly on $\gamma_q$,
and is nearly independent of $\lambda_s^{\rm eq}$ and $\gamma_s$ as
$\Lambda = (sdu)$ and $K^- = (s\bar{u})$.
In Fig.~\ref{datyld}
we show how the relative yield depends on $\gamma_q$ and $T$.
The $\Lambda$ yield we wish to consider
does not include weak decay feed from $\Xi$ but
it includes the electromagnetic decay of $\Sigma^0$ and the strong decays.
$K^-$ excludes feed-down from $\phi$, but includes $K^*$ and higher resonances.
It is important to exclude the  $\Xi$ and $\phi$ cascading in order to
eliminate the dependence on $\gamma_s$ and $\lambda_s^{\rm eq}$.
Fortunately, this is  experimentally feasible.   

A similar ratio, which is
experimentally easier to correct
for, is $\Xi/\phi$, also dependent on temperature and $\gamma_q$ only.  See
\cite{ourfluct2} for the equivalent discussion in
terms of $\Xi/\phi$.
%
%%%%%%%%%%%%%%%%%%%%%%%%%%
%%%%%%%%%%%%%%%%%%%%%
%

\begin{figure}[t]
\psfig{width=8.cm,clip=,figure=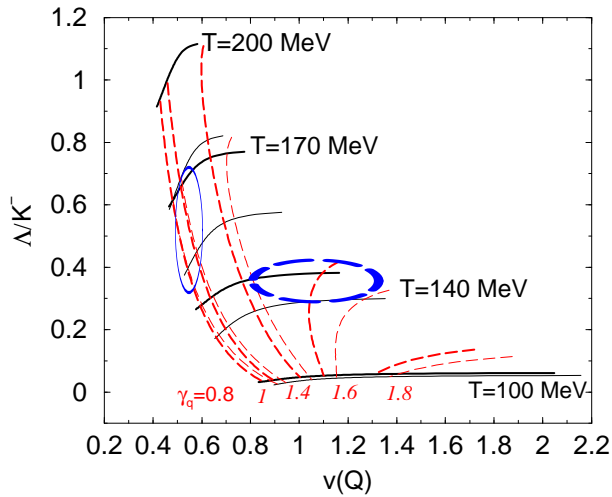}
\caption{(Color online)\label{datyldfluct}
 Particle   ratio   $\Lambda/K^-$  and particle fluctuation
$v(Q)$  plane: a point in plane
corresponds to a set of values  $\gamma_q, T$.
Black Lines correspond to results at fixed $T=200$ (top),
 170, 140, 100 MeV (bottom). The
red dashed lines are for  $\gamma_q=0.8,1,1.4,1.6,1.8$ from left to right.
Thick lines correspond to $\gamma_s=2.5$, thin lines correspond to
$\gamma_s=1$.
Ellipses (blue) indicate the expected result areas for
the equilibrium ($\gamma_q=1$, solid) and non-equilibrium
($\gamma_q\ne 1$, dashed) models.}
\end{figure}

We now combine   results in Figs.~\ref{datfluctboltz} and \ref{datyld}
into our main result Fig.~\ref{datyldfluct}.
Every point in this plane of $v(Q)$ and
$\Lambda/K^-$ corresponds to a specific set of $T$ and $\gamma_q$
as indicated by the grid. Note that
some domains in this plane are not allowed
since they lie in the region where the (generating, GC) partition function
cannot be defined.
The two highlighted regions indicate the
expected chemical equilibrium (solid line ellipse at small $v(Q)$,
corresponding to
$\gamma_q=1$ and $T=170$ MeV) and nonequilibrium
parameter domains (dashed line ellipse at larger $v(Q)$, corresponding to
$\gamma_q=1.62$ and $T=140$ MeV). When particle yields and
fluctuations are considered, the separation of these two
domains confirms that we have found  a sensitive method to determine
both $\gamma_q$ and $T$.
%
%%%%%%%%%%%%%%%%%%%%%%%%Figure 3
%%%%%%%%%%%%%%%%%%%%%
The results of having two extreme values,
$\gamma_s=1$ and $\gamma_s=2.5$, are also
shown in Fig.~\ref{datyldfluct}. The $\gamma_s$ values  corresponds to the
equilibrium \cite{bdm} and non-equilibrium \cite{Rafelski:2003ju} best fits.
Their difference, as seen in Fig.~\ref{datyldfluct}, is small and
well below the experimental error.
%As discussed, the reason is that in all models $v(Q)$ is controlled
%by non-strange particle fluctuations.
%The ratio $\Lambda/K^-$ is also independent
%of $\gamma_s$, since we have corrected for strangeness-changing
%$\Xi \rightarrow \Lambda$ and $\phi \rightarrow K$ decays.
The largest remaining systematic deviation is due to the baryon chemical
potential $e^{\mu_B/3 T} = \lambda_q^{\rm eq}$.
It's contribution to $v(Q)$ is negligible,
but this is not true for the case of $\Lambda/K^-$.
Generally the value of $\lambda_q^{\rm eq}$ is well determined
by  baryon to antibaryon yield ratios in a model independent way.

To transform the diagram in Fig.~\ref{datyldfluct} (or $\Xi/\phi$ in
\cite{ourfluct2})  to an
equivalent result applicable to lower reaction energy where
$\lambda_q^{\rm eq}$ is greater, one has to allow for this
change: We note that $\Lambda/K^-\propto (\lambda_q^{\rm eq})^3$,
and thus we need to multiply the
axis in Figs.~\ref{datyld} and \ref{datyldfluct} by
 $(\lambda_q^{\rm eq})^3/1.05^3$. One can actually use the
$\Lambda/K$ ratio in this. Since
${\Lambda K^+}/{\overline{\Lambda} K^-}\propto (\lambda_q^{\rm eq})^6$,
the axis rescaling would be done with
$({\Lambda K^+}/{\overline{\Lambda} K^-})^{1/2}/1.05^3$
($\Lambda,K$ corrected for $\Xi$ and $\phi$ feed-down).

\section{\label{secacceptance}Issues related to detector acceptance}
%%%%%%%%%%%%%%%%%%%%%%%%%%%%%%%%%%%%%%%%%%%%%%%%%%%%%%%%%%%%%%%%%%%%%

The main phenomenological issue that prevents the straight-forward extraction
of parameters from graphs such as Fig. \ref{datyldfluct} are effects relating
to the detector acceptance.
First of all, it has long been known that $v(Q)$ is not a ``robust''
observable, but in general depends on the detector's kinematic (rapidity and $p_T$) cuts.
This difficulty, however, can be lessened via mixed
event background subtraction.  It can be shown \cite{fluct6} that observables corrected
this way are  in certain limits ``robust'' w.r.t. kinematic cuts and detector response.

We have discussed how to
generalize the methods described in this paper to robust observables
elsewhere \cite{ourfluct1,ourfluct2,ourfluct3}, and hence will not dwell on
this topic, beyond noting that,
while diagrams such as Fig. \ref{datyldfluct} need to be re-thought since
dynamical observables generally also depend on the (average) system volume, the
{\em sensitivities} of the fluctuation and yield observables to the
statistical model parameters follow the pattern described by this paper.
Hence, generalizing the methods described by this paper to dynamical
observables (whether via fits, as was done in \cite{ourfluct3} or
three-dimensional diagrams), is not a difficult task.

An issue that needs to be addressed separately, however, is the acceptance
dependence of particle {\em correlations}.   If the detector's pseudo-rapidity
coverage is too large, than the small volume assumption
required for the Grand-Canonical ensemble becomes untenable, and long-range
correlations (such as global conservation laws) can modify fluctuations.
If the detector's pseudo-rapidity coverage is too small, correlations due to
resonance decays acquire a rapidity-dependent correction (which is {\em not}
eliminated by mixed-event subtraction since it
corrects {\em two-particle correlations}).   We will address these issues in
the next sub-sections.

\subsection{\label{secconserv}Influence of conservation laws on fluctuations}

If the detector can capture the full phase space of the system than, barring
dramatic departure from standard model physics, the net charge of the event can
not fluctuate.
More generally, if the phase space size of the detected system becomes
comparable to the total system size, observables will not anymore be given by
the Grand-Canonical ensemble.

If the system is a fluid (or in general not in {\em global} equilibrium) {\em
no} ensemble is expected to provide a good description of fluctuations beyond the small volume Grand Canonical limit, since
the observable region of phase space will include many locally equilibrated
volume elements exchanging energy and quantum numbers via hydrodynamic flow.
While yields could still be approximated by some ensemble, the long range
correlations and global non-equilibrium should break all simple scaling of
fluctuations with yields.

Hence, the configuration space coverage needed for a statistical description
needs to be appropriately small for the corrections to the GC ensemble to be
kept under control.

To investigate these corrections quantitatively, consider the
Taylor-expansion of the
entropy of the ``reservoir'': 
\begin{eqnarray}
\label{gcdef}
S(N_{\rm tot}-N) & \approx & S(N_{\rm tot})
-N \left.  \frac{\partial S}{\partial N} \right|_{N_{\rm tot}} 
\nonumber \\ 
& & {} 
+   \frac{1}{2} N^2 \left.
\frac{\partial^2 S}{\partial N^2} \right|_{N_{\rm tot}} + ...
\end{eqnarray}
where $N_{\rm tot}$ is the total number of particles in the reservoir and 
the small subsystem, and $N$ is the number of particles in the subsystem.
The first and second terms result in the usual Grand-Canonical
ensemble result \cite{huang} through the identification of the equilibrium chemical potential
$\mu = -T (\partial S/\partial N)$.

The third term gives the first correction;
The Grand-Canonical
ensemble is therefore a valid approximation when
\begin{equation}
\label{gccorr}
\zeta_{GC} = \frac{\ave{N}}{2}
\frac{({\partial^2 S}/{\partial N^2})_{N_{\rm tot}}}
{({\partial S}/{\partial N})_{N_{\rm tot}}}
\ll 1
\end{equation}
This quantity can be easily related to more common thermodynamic quantities
\begin{equation}
\zeta_{GC} = \frac{1}{2} \frac{T \ave{N}}{\mu}{k_{Vtot}} \end{equation}
where $\ave{N}$ is the average multiplicity of the {\em observed volume} and
$k_{Vtot}$ is the susceptibility of the {\em total volume}.
For the relativistic ideal gas, this is given by
\begin{equation}
\zeta_{GC}=  \frac{V}{2 V_{tot}} \ \left[ \frac{\sum_{n=0}^{\infty}
\lambda^n m^2 T K_2 \left( \frac{n m}{T} \right)}{ \ln \lambda
\sum_{n=0}^{\infty} \lambda^n m^2 \frac{T}{n} K_2 \left( \frac{n m}{T} \right)}
\right]
\end{equation}
and, as shown in section \ref{obschoice}
\[\ \frac{V}{V_{tot}}=\frac{\Delta \eta}{(\Delta \eta)_{tot}} \]
where $\Delta \eta$ is the detector's (pseudo)rapidity coverage and 
$(\Delta \eta)_{tot}$ is the system's rapidity interval.

Thus, we discover that the larger the susceptibility is, the smaller the system
size $V/V_{\rm tot}$ has to be for the Grand-Canonical limit to hold.

In fact, the physics determining the departure from this limit is {\em
precisely the same} as the physics determining the divergence of fluctuations
within an over-saturated pion gas.
This is unsurprising, since over-saturation is argued for as a signature of a
phase transition, and in finite systems undergoing phase transitions it is the
finite size of the system
that gives a cut-off for fluctuations.

The pion chemical potential of the system created at RHIC, however, is kept
below divergence, so it is hoped that one unit of rapidity, corresponding to
$V/(2V_{\rm tot}) \sim 7 \%$, provides a safe limit for the 
Grand Canonical ensemble.
In such a small rapidity interval, 
however, correlations due to resonances need to be
suitably accounted for.  The next
sub-section shows how to do that.

\subsection{\label{correso}Disappearance of resonance correlations at small
$\Delta \eta$}
If charge fluctuations are calculated {\em after} all resonances have decayed,
then Eq. \ref{eq:DQ2} becomes
\begin{equation}
\ave{(\Delta Q)^2} = \ave{(\Delta N_+)^2} + \ave{(\Delta N_-)^2} - 2
\ave{\Delta N_+ \Delta N_-}
\label{fluctdefcorrel1}
\end{equation}
where the last term accounts for unlike-sign charge 
correlations coming from the decay of neutral resonances.
For a conserved charge, and full acceptance of all resonances, this expression
is equivalent to Eq.(\ref{eq:DQ2}), with the
correlation term exactly balancing out the amplification of resonance abundance
fluctuations through the greater multiplicity of resonance decay products.
within a hadron gas the correlation term will be given by decays of the
resonance $j$ into $N_+$ and $N_-$
\begin{equation}
\label{correctioncorr}
 \ave{\Delta N_+ \Delta N_-} = \sum_j  b_{j\rightarrow + -} \ave{N_j}
\end{equation}
while the fluctuation of each stable $N_{\pm}$ has to be augmented by
contributions to it from resonance decays \cite{fluct1}
\begin{eqnarray}
\label{correctionres}
  \ave{(\Delta N_{\pm})^2}&=&\sum_i  \ave{(\Delta N_{\pm})^2}_{i}+\\
&&\hspace*{-2cm}+ \left( \sum_j b_{j \rightarrow i}(1-b_{j \rightarrow i})
\ave{N_j}
  + b_{j \rightarrow i}^2 \ave{(\Delta N_j)^2} \right)
\nonumber
\end{eqnarray}
For a finite acceptance window in general not all resonances
produced can be reconstructed, even if the efficiency of the
detector were 100\%. Hence these contributions must be weighted
with acceptance weight factors, and this applies here in particular
to the limited rapidity acceptance.
For a neutral resonance $j$ decaying
into $n_+$ positive particles and $n_-$ negative particles,
three such coefficients are needed:   

Two will be the fractions of the positively charged and
the negatively charged decay products which land in
the acceptance window, and the third
will give the fraction of
the $+-$ {\em pairs}
that will land in the window.
These coefficients will 
modify
the branching ratios $b_{j \rightarrow i}$
in Eq.(\ref{correctionres}) and $b_{j\rightarrow +-}$
in Eq.(\ref{correctioncorr}).

\begin{figure*}[!tb]
\psfig{width=8.cm,clip=,figure=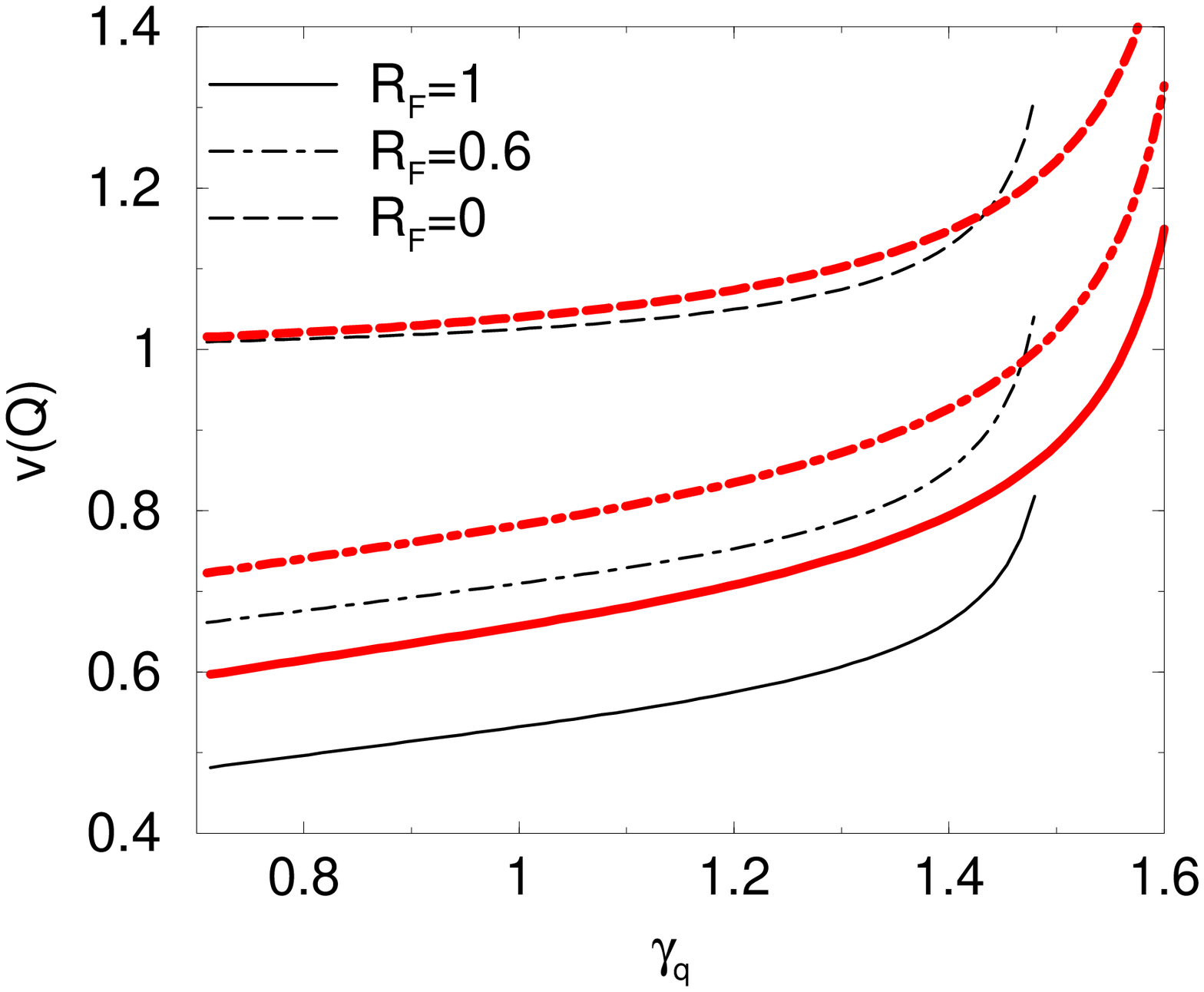}
\psfig{width=8.cm,clip=,figure=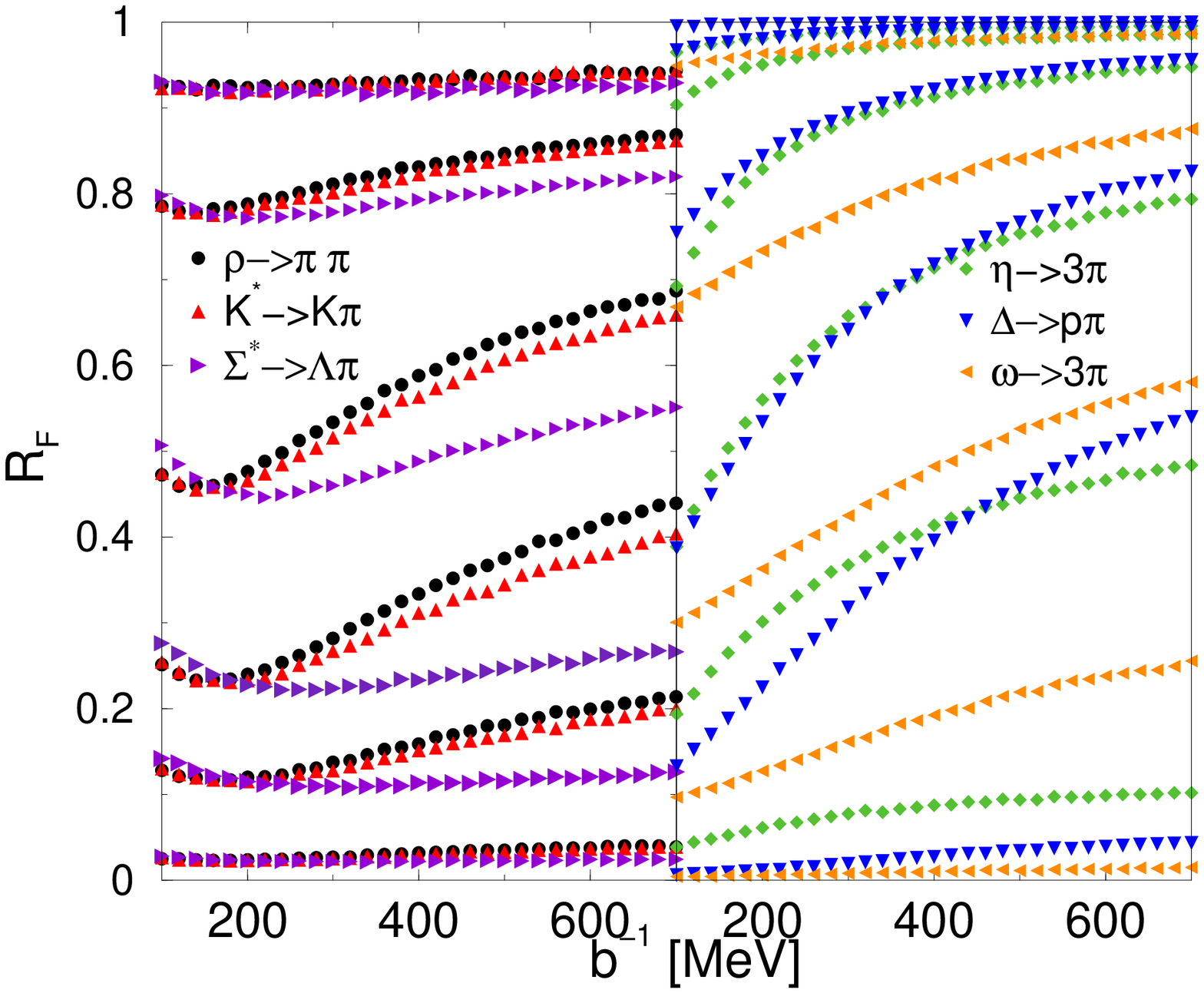}
\caption{(Color online) \label{expcorrel}
Left: Sensitivity of the charge fluctuation measure on $R_F$, the fraction of
resonance decay products which remains in the detector acceptance
window ({\it c.f.}\ Eq.(\ref{fluctdefcorrel2})).
Thin black lines denote $T=170$ MeV, thick
red lines $T=140$ MeV.  Right: Acceptance fraction for different resonance
decays as a function of the inverse slope $b$ ({\it c.f.}\ 
Eq.(\ref{slope})) and the
detector pseudo-rapidity acceptance $\Delta \eta$ (NB: $\eta$ in this context
means the pseudo-rapidity.  Not to be confused with the decay of the $\eta$
particle, shown on the right panel of the image).    Acceptance regions of
$\Delta \eta=6,4,2,1,0.5,0.1$ are considered, top to bottom in descending order
}
\end{figure*}

If boost-invariance is a good symmetry, the first two coefficients can be fixed
to unity, since particles coming \textit{out} of the acceptance region
are exactly balanced by particles coming \textit{in}.   However,  this is not
true for the number of detectable pairs.
If the resonance is out of the detector's acceptance window it is impossible
for {\em all} of it's decay products to be in a window.  Hence, 
Eq.(\ref{fluctdefcorrel1}) will have to include a
term giving the percentage of resonances whose decay products are both within
the detector's acceptance region.
\begin{eqnarray}
\ave{(\Delta Q)^2} &=& \ave{(\Delta N_+)^2} + \ave{(\Delta N_-)^2} \nonumber \\
&-& 2 R_F(T,\Delta y)  \ave{\Delta N_+ \Delta N_-}
\label{fluctdefcorrel2}
\end{eqnarray}
The  dependence of the observed fluctuations on $R_F$ is shown in Fig.
\ref{expcorrel}, left panel.

 We note two effects not considered here and believed to be unimportant:\\
1) the rescattering after formation is unlikely to alter $R_F$, since the
typical
momentum exchange in each collision the exchanged momentum
$\ave{q} \sim T/3$ tends to be considerably
softer
than what is required to bring particles outside the acceptance region (in most
decays, the characteristic
momentum of the decay products in a resonance's rest frame $p^*$ tends to be
significantly
larger than this value);\\
2) The higher-momentum pseudo-elastic ``regeneration'' processes, where
detectable
resonances would be created, are also unlikely to modify $R_F$ since, by local
thermal
equilibrium, two particles coming into the acceptance region through
kinematically allowed
pseudo-elastic interactions will be balanced out by two particles originally in
the
acceptance region which come out as a result of the re-interaction.\\
Thus, a measurement of fluctuations can still be relied upon to gauge the
number of
resonances present at {\em chemical} freeze-out.   This underscores the
importance of
fluctuations as a probe for freeze-out
dynamics.

We now obtain $R_F$ for a azimuthally symmetric perfect detector having a
pseudo-rapidity
coverage $\Delta \eta$.  We shall follow the formalism in \cite{Ani85}
to relate the resonance's
rest frame (denoted by $*$) to the lab frame.

For both particles $+$ and $-$ to be within
the detector's acceptance region, $-\Delta \eta/2 < \eta_{+},\eta_{-}< \Delta
\eta/2$ where
\begin{equation}
\label{pseudorap}
\eta_{\pm}= \frac{1}{2} \log \left( \frac{\sqrt{E_{\pm}^2 - m_{\pm}^2} -
p_{L\pm}}{\sqrt{E_{\pm}^2 - m_{\pm}^2} + p_{L\pm}} \right) = \ln \left[  \cot
\left( \frac{\theta_{\pm}}{2} \right) \right]
\end{equation}
If all angular dependence in the resonance's decay matrix elements is neglected
(a valid approximation if many resonances are produced, with an approximately
azimuthally invariant distribution)the fraction of detectable $+-$ pairs will
then be simply given by a phase space integral \begin{equation}
\Omega_{+-} (\eta_R,p_{TR}) = \int  \frac{d^3 p^*_+}{E^*_+} \frac{d^3
p^*_-}{E^*_-} \prod_i \frac{d^3 p^*_i}{E^*_i} \Theta_{+-}
 \label{correlpt}
\end{equation}
where:
\[\  \Theta_{+-} = \Theta  \left[\eta_{+}- \frac{\Delta \eta}{2} \right]
    \Theta \left[\eta_{+}+ \frac{\Delta \eta}{2} \right]  \Theta \left[\eta_{-}- \frac{\Delta \eta}{2} \right]
    \Theta \left[\eta_{-}+ \frac{\Delta \eta}{2} \right]  \]
and the function $\Theta(z)$ is the usual step function
\[\
\begin{array}{cc}
\Theta(z)=0 & z<0\\
\Theta(z)=1 & z>0\\
\end{array}   \]
Now, for two body decays this reduces to
\begin{equation}
\Omega_{+-} (\eta_R,p_{TR}) = \frac{1}{4 \pi} \int_0^{2 \pi} d \phi \int_0^1 d
\left( \frac{p_L^*}{p^*} \right) \Theta_{+-}
 \label{2bodypt}
\end{equation}
while for three body decays we use the Monte-Carlo routine MAMBO \cite{mambo}
to generate points in phase space.

%%%%%%%%%%%%%%%%%%%%%%%%%%
%%%%%%%%%%%%%%%%%%%%%

To calculate $\eta_+$ and $\eta_-$ from the resonance rest frame kinematic
variables we Lorentz-transform to the lab frame, and get \cite{Ani85}
\begin{eqnarray}
\label{plpm}
p_{L\pm} = \pm p^*_{L \pm} + \frac{p_{LR}}{m_R} \left( E^*_{\pm} +
\frac{\vec{p^*}.\vec{p_R}}{E_R+m_R}  \right)\\
\label{ptpm}
p_{T\pm} = \pm p^*_{T \pm} + \frac{p_{TR}}{m_R} \left( E^*_{\pm} +
\frac{\vec{p^*}.\vec{p_R}}{E_R+m_R}  \right)
\end{eqnarray}
To get an over-all fraction of accepted resonances which will enter 
Eq.(\ref{fluctdefcorrel2})
, one has to convolute Eq.(\ref{correlpt}) with a resonance distribution
function in momentum space
\begin{equation}
\label{eqrf}
R_F = \int_0^{\infty} d p_{TR} \int_{-\Delta \eta/2}^{\Delta \eta/2} d \eta_R
P(\eta_R,p_{TR}) \Omega_{+-} ( \eta_R,p_{TR})
\end{equation}
where $P(\eta_R,p_{TR})$ is a suitable distribution function for resonances
\textit{normalized to unity}.  A suitable function in the low energy region at
mid-rapidity is
\begin{equation}
\label{slope}
P(\eta_R,p_{TR}) = \frac{m_{TR}^\alpha e^{- b m_{TR}}}{\Delta \eta_R
\int_m^\infty d m_{TR} m_{TR}^\alpha e^{-b m_{TR}}}
\end{equation}

We have performed this integral using a Monte-Carlo method.  The result is
shown in the right panel of Fig. \ref{expcorrel}.
We note that the most abundant resonance decays for charge fluctuations do not
depend strongly on the
inverse slope parameter $b^{-1}$:  Going from $b^{-1}=200$ MeV to $b^{-1}=300$
MeV while staying in the same rapidity bin changes the $\rho \rightarrow \pi
\pi$ correction by at most 5 $\%$, and the less abundant but more sensitive
$\eta \rightarrow \pi^+ \pi^- \pi^0$ correction by no more than 20$\%$.

Thus,  $\Delta \eta$ should be 
as small as possible, statistics permitting, due to
the not easily controllable corrections described in section \ref{secconserv}.
A subsequent SHM analysis of the experimental data can than calculate
$R_F$ for each resonance decay important for charge fluctuations.
Hence, a $v(Q)$ , properly corrected for experimental acceptance,
can be computed from SHM parameters via Eqs.(\ref{vqdef}) and 
(\ref{fluctdefcorrel2}), and fed into
Fig. \ref{datfluctboltz} and similar
figures or fits \cite{ourfluct1,ourfluct2,ourfluct3}.
The computational tools needed to perform such an analysis have been
published  separately as open-source software \cite{share2}.

It is important to underline that to perform this analysis it is not
necessary to understand the full freeze-out dynamics of the fireball (local
temperature, flow field, hadronization hypersurface).   It is enough to have a
sensible parametrization of $b^{-1}$ in terms of particle mass.  This function
is commonly obtained from particle spectra at {\em thermal} freeze-out
\cite{slopemass}, and is approximately linear in particle mass.
The question is weather we can extrapolate $b^{-1}$ to  {\em chemical
freeze-out} conditions with enough precision 
in a model-independent way.    The relatively mild dependence of $R_F$ on
$b^{-1}$, together with the fact that
hadronic re-interaction decreases the temperature and increases the flow and
the high viscosity of the hadron gas \cite{hadvisc} makes us confident that
we can do it.

\section{Summary and conclusions}
%%%%%%%%%%%%%%%%%%%%%%%%%%%%%%%%%%%%%%%
We have studied in this work how a simultaneous
measurement of charge fluctuations and a ratio such as $\Lambda/K^-$
 can  differentiate
between  chemical equilibrium and non-equilibrium
freeze-out, and to constrain the magnitude of the deviation from
equilibrium as well as the freeze-out temperature.  Our results show that it is
possible to   distinguish
the chemical equilibrium freeze-out condition
$\gamma_q=1$  \cite{Rafelski:2004dp} with  $T=170$ MeV \cite{bdm})
from the chemical non-equilibrium freeze-out condition
$\gamma_q=1.6$ \cite{Rafelski:2003ju,Rafelski:2004dp}. This is mainly due to
the increase in the
fluctuations inherent to an oversaturated Bose gas, see Eq.(\ref{divergence}).

We have further discussed the dependence of two-particle correlations on the
detector acceptance region, and have shown that it can be
calculated to a reasonable precision in a model-independent way.  The ``right''
experimental detector acceptance for a detailed study of
fluctuations, therefore, is one that is appropriately small yet sizable to
ensure the appropriate ensemble under study is Grand-Canonical, provided that
acceptance corrections to resonance decays are properly taken into
account using the methods described in section \ref{correso}.      Quantitative
corrections to Grand Canonical yield/fluctuation relations
for the best fit parameters can be estimated quantitatively via 
Eq.(\ref{gccorr})

Provided the detector acceptance region for a given fluctuation measurement
is published, Eq.(\ref{eqrf})
can be used to calculate a correction coefficient $R_F$ to the
$\ave{N_+ N_-}$ correlation for each decay of a neutral resonance.
Using a calculated $R_F$ for each resonance decay, together with the
statistical model parameters, the charge fluctuation variable $v(Q)$ can be
calculated from
Eqs.(\ref{vqdef}) and  (\ref{fluctdefcorrel2}).
This $v(Q)$ will still retain
the sensitivities to temperature and $\gamma_q$ demonstrated in
section \ref{noneq}, since $\gamma_q$ impacts the primordial fluctuation terms rather than the correlation.   It can therefore be used, together with a measurement
such as $\Lambda/K^-$ as in Fig. \ref{datyldfluct}, or within a fit
as in \cite{ourfluct1,ourfluct2,ourfluct3}, to test the validity of the
statistical model, unambiguously constrain its parameters, and differentiate
between the high-temperature equilibrium and supercooled over-saturated
freeze-out scenarios.

It is our intent to perform a complete data analysis as outlined here,
including consideration of acceptance corrections and of resonance decays,
once final RHIC fluctuation data becomes available.

%%%%%%%%%%%%%%%%%%%%%%%%%%%%%%%%%%%%%%%%%
\subsection{Acknowledgments}
GT thanks C. Gale, L. Shi,
V. Topor Pop, A. Bourque, Wojciech Broniowski, Wojciech Florkowski and Mark Gorenstein for stimulating discussions and the Tomlinson foundation  for support.
S.J.~thanks RIKEN BNL Center
and U.S. Department of Energy [DE-AC02-98CH10886] for
providing facilities essential for the completion of this work.
Work supported in part by  grants from
the U.S. Department of Energy  (J.R. by DE-FG02-04ER41318),
the Natural Sciences and Engineering research
council of Canada, the Fonds Nature et Technologies of Quebec.
%%%%%%%%%%%%%%%%%%%%%%%%%%%%%%%%%%%%%%%%%
%\vskip 0.3cm
%%%%%%%%%%%%%%%%%%%%%%%%%%%%%%%%%%%%%%%%%
%\begin{references}

\end{document}